\documentclass[aps,prb,twocolumn,showpacs,superscriptaddress]{revtex4} %
\usepackage{epsfig,amsmath,amssymb}
\usepackage{graphicx}
\usepackage[english]{babel}

\begin{document}

\title{Quantum superposition of three macroscopic states \\ and superconducting qutrit detector} %
\author{V.I. Shnyrkov}
\email{shnyrkov@ilt.kharkov.ua} %
\affiliation{B. Verkin Institute for Low Temperature Physics and
Engineering, National Academy of Sciences of Ukraine, Lenin
Ave. 47, Kharkov 61103, Ukraine}%
\author{A.A. Soroka}
\affiliation{National Science Center ``Kharkov Institute of Physics
and Technology'', Akhiezer Institute for Theoretical Physics,
Akademicheskaya St. 1, Kharkov 61108, Ukraine}
\author{O.G. Turutanov}
\affiliation{B. Verkin Institute for Low Temperature Physics and
Engineering, National Academy of Sciences of Ukraine, Lenin Ave. 47,
Kharkov 61103, Ukraine}%

\begin{abstract}

Superconducting quantum coherent circuits have opened up a novel
area of fundamental low-temperature science since they could
potentially be the element base for future quantum computers. Here
we report a quasi-three-level coherent system, the so-called
superconducting qutrit, which has some advantages over a two-level
information cell (qubit), and is based on the qutrit readout circuit
intended to measure individually the states of each qubit in a
quantum computer. The designed and implemented radio-frequency
superconducting qutrit detector (rf SQUTRID) with atomic-size
ScS-type contact utilizes the coherent-state superposition in the
three-well potential with energy splitting $\Delta E_{01} /k_{B}
\approx 1.5\,$K at the 30th quantized energy level with good
isolation from the electromagnetic environment. The reason why large
values of $\Delta E_{01}$ (and thus using atomic-size Nb-Nb contact)
are required is to ensure an adiabatic limit for the quantum
dynamics of magnetic flux in the rf SQUTRID.
\end{abstract}


\maketitle

\section{Introduction}

The phenomenon of the superposition of states of a macroscopic
object predicted for a superconducting quantum interferometer device
(SQUID) in the low-dissipation limit\cite{1,2,3} was revealed in
spectroscopic experiments.\cite{4,5} Although much progress has been
made in demonstrating the coherent quantum behavior of various
systems with Josephson junctions,\cite{6} there has not been an
experimental presentation of a readout device based on the quantum
superposition of macroscopically distinct states in flux qubits. The
dependence of this fundamental property of quantum mechanics, the
superposition of states, on the symmetry of the potential energy in
flux qubits can be taken as a basis for creating a radio-frequency
superconducting qubit detector (rf SQUBID) similar to the manner of
how the Josephson current-phase relation is used in building rf
SQUIDs.\cite{7} This device would be a principal element in quantum
readout circuits meant for weak continuous measurement of states of
the flux qubits incorporated in the quantum computer architecture.

However, there is a pitfall on this way.  In the literature, the
flux qubit (a superconducting ring closed by a
superconductor-insulator-superconductor (SIS) junction) is
described as the superposition of the two states %
$\left|\Psi \right\rangle = %
c_{1}(t)\left|\Psi _{1}\right\rangle + c_{2}(t)\left|\Psi _{2}\right\rangle$ %
that appear in this quantum system with double-well potential. If
the tunneling amplitude is large (in other words, if coefficients
$c_{1}$ and $c_{2}$ vary quickly enough), the system behavior
becomes adiabatic when changing the external magnetic flux; that is,
it can be considered in terms of quasi-stationary superposition
levels. In this case, it becomes impossible to distinguish between
the experimental characteristics of a common classical SQUID in a
nonhysteretic regime and a double-well SQUBID. Both devices behave
as parametric inductances (Josephson inductance for SQUID and
quantum inductance for SQUBID), with both inductances being
maximized at the same external flux, $\Phi_{e}=(n+1/2)\Phi_{0}$
(where $n$ is an integer and $\Phi_{0}=h/2e$ is the superconducting
flux quantum), so that some additional evidence is required for
establishing the quantum nature of the object under study. Unlike
this situation, with the superposition of three classically
separated states in the superconducting ring, the characteristics of
a radio-frequency superconducting qutrit detector (rf SQUTRID) will
possess essential distinctions\cite{8}, allowing one to state
definitely their quantum origin. Particularly, the quantum
inductance extrema in the qutrit should be observed at external
magnetic flux $\Phi_{e}=n\Phi_{0}$.

Here we present experimental evidence and a theoretical analysis
for the fact that a rf SQUID with atomic-size ScS contact %
can be put into superposition of three distinct states with a
macroscopically large time of energy relaxation to lower levels and
thus be turned into a rf SQUTRID. We explore the voltage-current and
voltage-flux (signal) characteristics of this new device for the
first time. Note that in the qutrit we study, the superposition
states are formed due to the fast tunneling of flux through
potential barriers in the triple-well symmetrical potential in the
phase space and the removal of the degeneracy of states of equal
energy in each well, unlike in Ref.\,\onlinecite{QT}, where the
transmon with its three lowest energy levels was used as a qutrit
whose superposition states were prepared by exciting the transmon
from the base energy level to two higher ones.

A substantial difference between the Josephson properties of ScS and
SIS contacts at low temperatures has been predicted\cite{QPC1,CPC,
QPC2,10} in the microscopic theory of superconducting weak links.
The singular potential corresponds to the case of a ``clean'' ScS
contact with a dimension $d$ much smaller than the superconducting
coherence length $\xi_0$ and the electron elastic mean free path
$\ell$ (the so-called ballistic regime $d\ll \ell$). %
This fact leads to some peculiarities in the macroscopic quantum
tunneling\cite{11,12} and the quantum coherence of magnetic flux
states in a superconducting ring closed by ScS contact. The major
ones are an appreciable increase in the qutrit key performance
parameter, the splitting of degenerate energy levels in separate
potential wells,\cite{8} and the emergence of high nonlinearity in
the resulting qutrit superpositional levels that can be used in the
rf SQUTRID based on rf SQUID circuitry (Fig.\,1). %

\begin{figure}[t!]
\centering %
\includegraphics[width = 1.0\columnwidth]{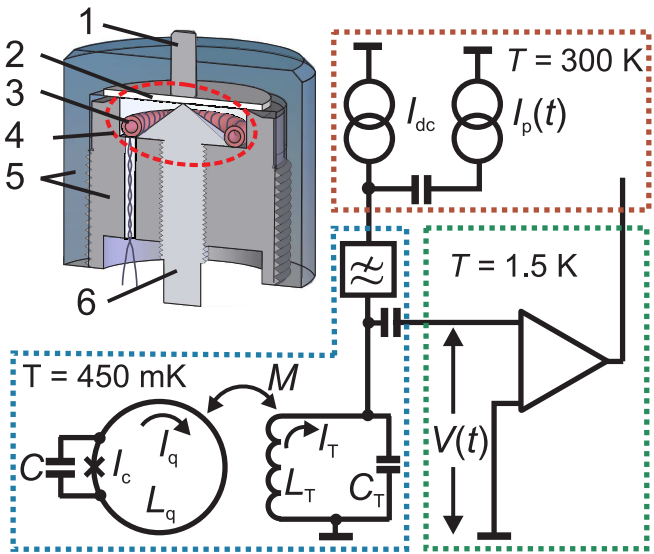}
\caption{\label{fig01} %
(Color online) Circuit diagram of the SQUTRID. Experimental set-up
is characterized by the following parameters: %
$L_{T}=1.2\,\mu$H, $C_{T}=630$\,pF, $I_{P}(t)=I_{0}\cos\omega t$,
$\omega_T/2\pi=5.794$\,MHz, $Q=302$,
$M=k\sqrt{L_{q}L_{T}}=1.52$\,nH, and $L_{q}=0.3$\,nH; %
the qutrit values $I_{c}$ and $C$ are discussed in the text. The
inset shows the fully niobium 3D toroidal construction of the qutrit cell, %
where 1 is the pusher, 2 is the membrane, 3 is a part of the $L_{T}$ coil, %
4 is the toroidal quantization loop (cavity), 5 is the body, and 6 is the needle. %
}%
\end{figure}

\section{The SQUTRID model}
Currently, two types of point contacts are distinguished, depending
on the ratio between the contact dimension $d$ and the electron wave
length $\lambda_F=h/p_F$: $d\gg \lambda_F$ for a \textit{classical}
point contact and $d\sim \lambda_F$ for a \textit{quantum} point
contact.\cite{QPC1} In metals, actually, a quantum point contact is
necessarily of atomic dimensions, as the electron wave length is of
the same order of magnitude as the atomic separation. For both
classical\cite{CPC} and quantum\cite{QPC2} ScS point contact with
the critical current $I_{c}$, at $T=0$ the current-phase relation
reads
\begin{equation} \label{EQ00}
\begin{array}{c}
\displaystyle{ %
I_s(\varphi) = I_c\sin\frac{\varphi}{2}\,\textrm{sgn}[\cos\frac{\varphi}{2}], %
\,\,  I_{c} = \frac{\pi\Delta_{0}}{e R_{N}},
} %
\end{array}
\end{equation}
where $\Delta_{0}$ is the superconducting energy gap and
$R_N$ is the normal-state resistance of ScS contact. %
The critical current of the atomic-size (quantum) ScS contact was
predicted\cite{QPC2} to be quantized (as a consequence of the
quantization of the contact conductance $R_N^{-1}$, in units of
$G_0=2e^2/h$), $\displaystyle{ I_{c}= N(e\Delta_0/\hbar)}$, %
which was observed experimentally.\cite{QPC1} %
From (\ref{EQ00}) one gets the Josephson coupling energy of ScS
contact in the form $U_J=-(I_c\Phi_0/\pi)|\cos(\varphi/2)|$. %

To develop the SQUTRID model in a zero-temperature approximation, we
take the quantum Hamiltonian in the flux representation of the
superconducting loop (Fig.\,1) of inductance $L_q$ closed by a clean
atomic-size ScS contact with critical current $I_{c}$ and
self-capacitance $C$ (so that parameter $g=E_J/E_C =\Phi_0I_c
C/(2\pi e^2) \gg 1$) in the form\cite{8,11,12,13}
\begin{equation} \label{EQ01}
\begin{array}{c}
\displaystyle{ %
\hat{H}_{q} =\frac{\hat{P}^{{\rm 2}} }{{\rm 2}M} +\hat{U}(f;f_e)= %
} \vspace{2mm}\\%
\displaystyle{ %
=-\frac{\hbar^{2}}{2M}\frac{\partial ^{2}}{\partial f^{2}}+ %
{\frac{\Phi_{0} I_{c}}{2\pi }} \left[-2\left| \cos( \pi f ) \right|+\frac{2\pi ^{2} (f-f_{e})^{2}}{\beta_{L} } \right], %
} \vspace{2mm}\\%
\displaystyle{ %
M=\Phi_{0}^{2}C, \quad \beta_{L}=\frac{2\pi L_qI_{c}}{\Phi_{0}}, %
}
\end{array}
\end{equation}
where $f= \Phi/\Phi_0$ and $f_{e}=\Phi_e/\Phi_0$ %
are the normalized internal magnetic flux $\Phi$ in the loop %
and external magnetic flux $\Phi_e$ applied to the loop. %
The quantum dynamical observable of the internal magnetic flux in
the loop is given by an operator of flux conjugated to an operator
of charge in the contact capacitance: $[\hat{\Phi},\hat{Q}]=-i\hbar$.\cite{13} %
The key feature of Hamiltonian (\ref{EQ01}) is its singular
potential $U(f;f_e)$ following from the nonsine current-phase
relation (\ref{EQ00}) for ScS contact. Note that a model with both
the potential attributed to ScS contact and the dissipation
vanishing at zero temperature can satisfactorily describe the
experiments on macroscopic quantum tunneling in a ring with a clean
ScS contact,\cite{12} as shown in Ref. \onlinecite{11}.

The solutions of the stationary Schr\"{o}dinger equation
\begin{equation} \label{EQ02}
  \hat{H}_{q}(f;f_{e})\,\Psi(f)=E(f_{e})\,\Psi(f)
\end{equation}
with Hamiltonian (\ref{EQ01}) yield wave functions $\Psi(f)$ and
energies $E(f_{e})$ of the stationary states of the superconducting
loop with ScS contact at a specified external magnetic flux $f_{e}$.
Let us consider the SQUTRID superconducting loop where a three-well
potential is formed. We get a series of states appeared during fast
(with rf generator rate $\omega$) increasing of external flux
$\Phi_{e}$ from 0 to $\Phi_{0}$ (Fig. 2) using parameters in Eq.
(\ref{EQ02}) close to our experimental values: $L_{q}=0.3$\,nH,
$C=4.36$\,fF, and $\beta_{L}= 4.0$ ($I_{c}\approx 4.4\,\mu$A). %
It is seen from these solutions that, with a change in external
flux, the initial state of a three-well symmetrical potential
localized in the central well at $\Phi_{e}=0$ [Fig.\,2(a)]
transforms through intermediate states [Fig.\,2(b) and 2(c)] into a
superposition state in a three-well symmetrical potential at
$\Phi_{e}=\Phi_0$ [Fig.\,2(d)]. Note that the energy exchange rate
between the two classically separated states in a two-well
symmetrical potential at $\Phi _{e}=\Phi_{0}/2$ is exponentially
small at the specified parameters, so the system state remains
localized in the starting potential well during the increase in
external flux toward the point $\Phi _{e}=\Phi_{0}$.
In this point, the superposition qutrit state %
${\left|\Psi_{S0}\right\rangle}=c_{1}{\left|\Psi_{1}\right\rangle}+ %
c_{2}{\left|\Psi_{2}\right\rangle}+c_{3}{\left|\Psi _{3}\right\rangle}$ %
of the wave functions of all three separate wells is formed in the
three-well symmetrical potential, similar to the formation of the
superposition qubit state in the two-well potential.

\begin{figure}[t!]
\centering %
\includegraphics[width = 1.00\columnwidth]{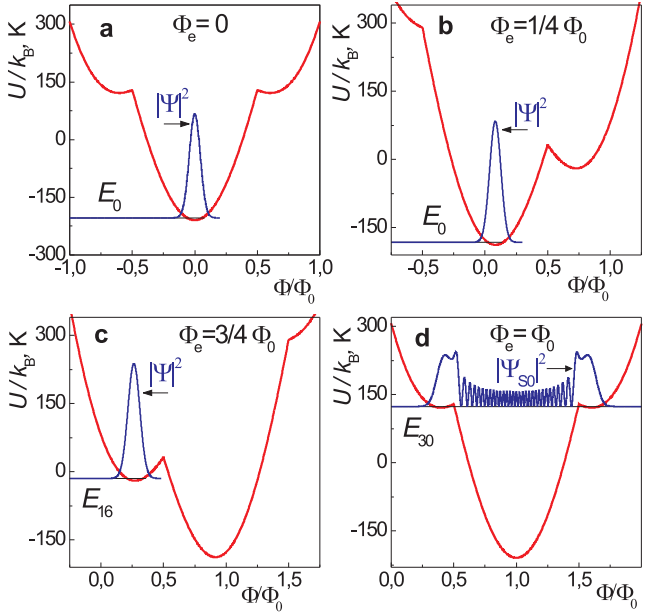}
\caption{\label{fig02} %
(Color online) Forming the base superposition SQUTRID state by rapid
change of external flux $\Phi_{e}$. The set of nonstationary
potential energies $U(f)/k_{B}$, in terms of temperature, and
squared wave functions $\left|\Psi(f)\right|^{2}$ are shown vs
normalized internal flux $f=\Phi/\Phi_{0}$ at various $\Phi_{e}$;
$\beta_{L}=4.0$. Subscript after $E$ refers to the energy level
number. (a) The SQUTRID is initially in its ground state; its wave
function is localized in the absolute minimum of the symmetrical
potential. (b) The potential becomes asymmetrical with two wells,
but the system is still in its ground state, and the wave function
is localized in the minimum of the deeper well. (c) The wave
function remains surprisingly localized in the same single well
since the potential changes too quickly in comparison with the
tunneling rate (the interwell barrier is rather high in this case).
(d) The base state $\left|\Psi_{S0}\right|^{2}$ of the three-well
superposition is formed in the potential symmetry point when the
barriers are small, and tunneling time is consequently short. The
relaxation time of this state turns out to be macroscopically large,
so that it becomes stable enough and ``latched'' for many further
cycles of $\Phi_{e}$.
}%
\end{figure}

Quantum coherence of the qutrit flux state
${\left|\Psi_{S0}\right\rangle}$ in a three-well potential manifests
itself as coherent oscillations of magnetic flux between all the
three potential wells due to its fast tunneling through the
potential barriers separating the central and the side wells.
Numerical analysis of Eq. (\ref{EQ02}) shows that, for resonant
tunneling in a three-well potential of the superconducting loop of
inductance $L_{q}=0.3$ closed by a clean ScS contact with $C=3 -
6\,$fF ($\beta_{L}=4.0$), the flux oscillation rate between wells
$\nu_{01}=\Delta E_{01}/h$ reaches $25 - 40$\,GHz. The magnetic
moment of the flux states corresponding to the side wells of the
three-well potential $\mu_s^{(1,3)}=I_sS
\simeq 10^{-11}$\,J/T\,$\simeq 10^{12} \mu_B$ %
(where $I_s\simeq 0.6\Phi_0/L_q \approx 4\,\mu$A is the supercurrent
in the side-well flux states, $S\simeq 2\!\times\! 10^{-6}\,$m$^2$
is the loop area, and $\mu_B=0.93\!\times\!10^{-23}$\,J/T is the
Bohr magneton), and the magnetic moment of the central flux state
$\mu_s^{(2)}=0$ (since the supercurrent is zero in this state). %
Thus, we have the coherent superposition of wave functions
corresponding to the {\it three distinct macroscopic} flux states in
the three-well symmetrical potential of the qutrit at
$\Phi_{e}=n\Phi_{0}$.

\begin{figure}[t!]
\centering %
\includegraphics[width = 1.0\columnwidth]{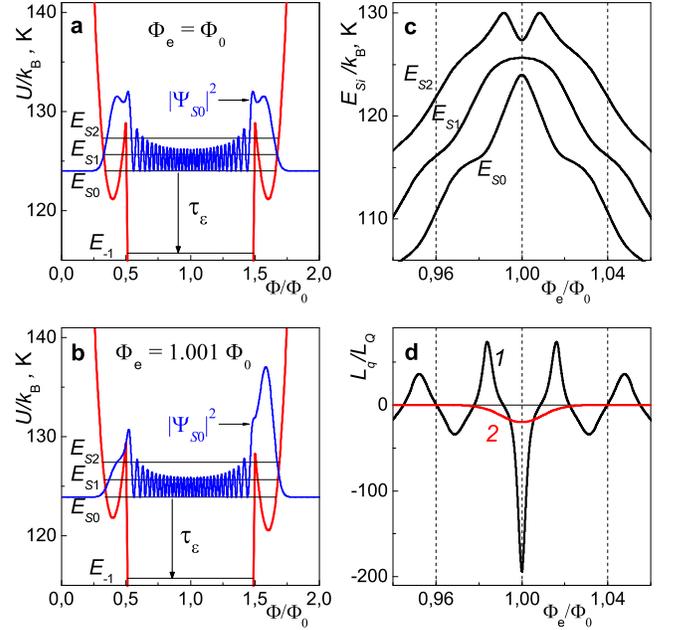}
\caption{\label{fig03} 
(Color online) Superposition of three states in the superconducting
loop closed by a clean ScS contact. Distributions of squared wave
functions $\left|\Psi_{S0}(f)\right|^{2} $ among the three potential
wells vs normalized internal flux $f=\Phi/\Phi_{0} $ for two close
values of external magnetic fluxes $\Phi_{e}$ corresponding to (a)
fully symmetrical and (b) slightly tilted potential $U(f)$;
$\beta_{L}=4.0$. Superposition energy levels are denoted $E_{S0},
E_{S1},$ and $E_{S2}$. The process of energy relaxation of the base
superposition state $\Psi_{S0}(f)$ from level $E_{S0}$ to level
$E_{-1}$ with characteristic time $\tau_{\varepsilon}$ is shown by
arrows. (c) Superposition energy levels $E_{S0}, E_{S1},$ and
$E_{S2}$ vs normalized external magnetic flux
$f_{e}=\Phi_{e}/\Phi_{0}$. (d) Normalized quantum inductance of the
base superposition level $E_{S0}$ vs $f_{e}$: curve 1 is zero-noise
$(L_{q}L_{Q}^{-1})(f_{e})$ and curve 2 is noise-affected
$(L_{q}L_{Q}^{-1})_{eff}(f_{e})$ averaged over low-frequency noise
with $\sigma =\sqrt{\left\langle \delta
f_{e}^{2}\right\rangle}=0.013$.
}%
\end{figure}

Let us refer to ${\left|\Psi_{S0}\right\rangle}$, which is the
three-well superposition state with minimum energy, as the ``base''
SQUTRID superposition state since adiabatic movement along the
respective energy level $E_{S0}$ plays the main role in our
experiments. State ${\left|\Psi_{S0}\right\rangle}$ in the potential
$U(f;f_{e})$ is characterized by quantum number $n\gg 1$ [$n=30$ for
the above-cited parameter values;  see Fig.\,2(d)]; i.e., the deep
central well ($\Delta U/k_B \simeq 300$\,K) contains a large number
of quantum levels. %
The time $\tau_{\varepsilon}$ of the energy relaxation of the base
SQUTRID state to underlying states in the central well must be
macroscopically large to enable measuring the superpositional
nonlinearity. This is achieved due to the design of the qutrit loop
in the form of a high-quality three-dimensional (3D) toroidal
superconducting cavity (see inset in Fig.\,1 and Sec.\,III), which
has no resonant modes with frequencies corresponding to the
frequencies (energies) of transitions from the base superpositional
level $E_{S0}$ to underlying energy levels.

Figure\,3(a) displays numeric solutions of Eq.\,(\ref{EQ02}) for the
squared absolute value of the wave function of the base
superposition state $|\Psi_{S0}(f)|^{2}$ and energy levels $E_{S0},
E_{S1}$, and $E_{S2}$ of the three superposition states as well. The
fact that splitting $\Delta
E_{01}/k_{B}=(E_{S1}-E_{S0})/k_{B}=1.65$\,K between the base and the
nearest superposition energy levels substantially exceeds the
environment temperature for the chosen parameters is key in the
present experiment ($\Delta E_{01}/ k_{B} T = 1.65/0.45 \approx 3.7
$). As a result, the broadening of level $E_{S0}$ because of
statistical
averaging over equilibrium density matrix %
$\delta E_{S0} =\Delta E_{01}/ [1+\exp(\Delta E_{01}/k_{B}T)]\ll
\Delta E_{01}$ is relatively small ($\delta E_{S0} /k_{B} \approx
0.04$\,K) and can be neglected in the analysis of the system energy
structure (zero-temperature approximation). Also, it is obvious that
the large tunnel splitting, which multiply overruns the temperature
value, guarantees good ``quantumness'' of the superconducting loop
closed by a clean ScS contact provided that the environment and
the measurement circuit noises are effectively suppressed. %
At the same time,  the base superposition level $E_{S0}$ is situated
far below the potential barrier top $U_b$, $(U_b-E_{S0})/k_B\approx
5$\,K$\,\,\gg T$, so that thermal transitions rate over the barrier
is vanishingly small, and the process of relaxation of the
metastable base superposition state into the deep central well can
be strongly limited for the chosen parameters since
$(E_{S0}-E_{-1})/k_{B}\approx 8$\,K.

Small adiabatic variations of the external flux relative to symmetry
points $\Phi_{e}=n\Phi_{0}$ lead to tilting the three-well
potential, with a change in the wave functions and the energy
levels. As seen from Fig.\,3b, when the flux shifts apart from the
symmetry point $\Phi_{e}=\Phi_{0}$, the wave packet of the base
state $|\Psi_{S0}(f)|^{2}$ partially transfers into the deeper side
potential well. A rearrangement of the superposition wave packets
and thus the formation of the superposition energy levels
$E_{S0}(f_{e}), E_{S1}(f_{e})$ and $E_{S2}(f_{e})$ [shown in
Fig.\,3(c)] occur over the
time span reciprocal to the flux tunneling rate, %
$t_{S}\approx\nu_{01}^{-1}$ . In experiment, we deal with relatively
low-frequency (adiabatic) processes, in which the time-dependent
superposition levels practically coincide with the stationary ones.

The superconducting qutrit detector is built on the basis of rf
SQUID circuitry. That is, the qutrit is inductively coupled to a
high-quality $L_TC_T$ superconducting tank (Fig.\,1) serving as
a linear classical detector ($\hbar\omega_T \ll k_BT$). %
Only a small fraction of the flux in the $L_TC_T$ tank ($M/L_T
\approx 0.0013$ in our experiment, where $L_T$ is the tank
inductance and $M$ is the mutual inductance between the qutrit and
the tank) is transferred to the qutrit, which allows making use of
the concept of weak continuous quantum measurements.\cite{WCM1,WCM2,WCM3} %
When the characteristic frequency of the classical detector
($L_TC_T$ tank) is much lower than that of the quantum object (in
our system the ratio $\omega/\nu_{01}\sim 10^{-3}$), the latter's
dynamics can be studied by means of quantum-mechanical equations
with the detector classical field treated as the external adiabatic
parameter.\cite{IMTt1,IMTt2,IMTt3} %
Within this semiclassical approach of treating the qutrit plus tank
system, one obtains the classical equation for the $L_{T}C_{T}$ tank %
containing the parametric \textit{quantum} inductance contribution
[see Eq.\,(\ref{EQ06})], instead of the parametric Josephson
inductance contribution probed in rf SQUID by means of the impedance
measurement technique (IMT). %
The IMT with a weak continuous quantum readout was successfully
applied to studying different types of
qubits.\cite{IMTe1,IMTe2,IMTe3}

For the $L_{T}C_{T}$ tank driven by rf-bias current $I_{0}\cos\omega
t$, the measured output functions are the amplitude of voltage
oscillations $V_T$ and the voltage-current phase shift $\alpha_{T}$. %
The equation for voltage $V(t)$ across the $L_{T}C_{T}$ tank having
the quality factor $Q=\omega_{T}R_{T}C_{T}$, %
with the current contribution due to the weakly coupled qutrit
$MI_{q0}/L_T$ taken into account, reads
\begin{equation} \label{EQ04}
C_{T}\dot{V}+\frac{V}{R_{T}}+\frac{1}{L_{T}}\!\int\! V dt+
\frac{MI_{q0}}{L_{T}}=I_{P}(t), \,I_{P}(t)=I_{0}\cos\omega t. %
\end{equation}
Here
\begin{equation} \label{EQ05}
I_{q0}(\Phi_{e})=\frac{\partial \langle \Psi_{S0}| \hat{H}_{q}|\Psi_{S0} \rangle }{\partial \Phi_{e} } = %
\frac{\partial E_{S0}(\Phi_{e})}{\partial \Phi_{e}}  %
\end{equation}
is the current circulating in the superconducting loop of the qutrit
in its base quantum superposition state, as a function of the
\textit{total} external flux $\Phi_{e}$. %

In further considerations it is convenient to use the function
$L_{Q}^{-1}(f_{e})$, which is called the reciprocal quantum
inductance, defined as
\begin{equation} \label{EQ03}
L_{Q}^{-1}(f_{e})= %
\frac{\partial I_{q0}(\Phi _{e})}{\partial \Phi_{e}}= %
\frac{\partial^{2} E_{S0}(\Phi _{e})}{\partial \Phi_{e}^{2}} = %
\frac{1}{\Phi_{0}^{2}}\frac{\partial^{2} E_{S0}(f_{e})}{\partial f_{e}^{2}}. %
\end{equation}
This function, being, in fact, the local curvature of energy level
$E_{S0}(f_{e})$, describes the nonlinear properties of the qutrit in
the base quantum superposition state at small variations of the
external magnetic flux. At the same time, it characterizes the
parametric inductance inserted in the $L_{T}C_{T}$ tank due to a
weakly coupled \textit{quantum} device.

Considering the emf induced in the qutrit loop
$\dot{\Phi}_{e}=M\dot{I}_{L}=MV(t)/L_{T}$,
we obtain %
\begin{equation} \label{EQ06}
\begin{array}{c}
\displaystyle{ %
\ddot{V}+\omega_{T}^{2} V= f(V,\dot{V},t)\,, %
} \vspace{2mm}\\%
\displaystyle{ %
f(V,\dot{V},t)= %
-k^{2}\,L_{q}L_Q^{-1}\![\Phi_{e}(t)]\,\omega_{T}^{2} V %
-\frac{\omega_{T}}{Q}\dot{V} + \frac{1}{C_{T}}\dot{I}_{P}\,, %
} \vspace{2mm}\\%
\displaystyle{ %
\Phi_{e}(t)= \Phi_{dc}+\Phi_{ac}(t)= \Phi_{dc}+\frac{M}{L_{T}}\int\!
V(t) dt \,,
} %
\end{array}
\end{equation}
where $\Phi_{dc}$ is the \textit{direct} (low-frequency signal)
external flux biasing the qutrit loop and $\Phi_{ac}(t)$ is the
\textit{alternating}
external flux applied to the loop due to the tank flux oscillations. %
Thus, the strongly nonlinear reciprocal quantum inductance function
$L_{Q}^{-1}(f_{e})$ [Eq.\,(\ref{EQ03})], characterizing the
curvature of the qutrit base superposition energy level, will
determine the solution
of Eq.\,(\ref{EQ06}). %
If the condition $f(V,\dot{V},t)\ll\omega_{T}^{2}\,V$
is fulfilled, which is valid when %
$Q\gg 1,\,I_{0}\ll\omega_{T}C_{T}V_{T},\,k^{2}L_{q}L_{Q}^{-1}\ll 1,$ %
the Krylov-Bogolubov method for solving weakly nonlinear
equations\cite{14} can be applied to solve Eq.\,(\ref{EQ06}).
Substituting voltage $V(t)$ in the form
\begin{equation} \label{EQ07}
V(t)=V_{T}(t)\cos[\omega t+\alpha_{T}(t)],
\end{equation}
where $V_{T}(t)$ and $\alpha_{T}(t)$ are slowly varying functions
(with small relative variation over the oscillation period $T=2\pi /\omega_{T}$), %
we get abridged equations for $\dot{V}_{T}(t),\,\dot{\alpha}_{T}(t)$, %
and the equations for the voltage amplitude and phase shift of
steady-state [$\dot{V}_{T}(t)=0,\,\dot{\alpha}_{T}(t)=0$] %
oscillations in the $L_{T}C_{T}$ tank: 
\begin{equation} \label{EQ08}
\begin{array}{c} %
\displaystyle{ %
V_{T}=\frac{\omega_{T}L_{T}Q\, I_{0}}{\sqrt{1+4\xi(V_{T},\Phi_{dc})^{2}Q^{2}}}, \,\, %
\tan\alpha_{T}=-Q(1+\xi_{0})\xi, %
} \vspace{2mm}\\ %
\displaystyle{ %
\xi =\xi_{0}-\frac{k^{2}}{2\pi}\!\int\limits_{0}^{2\pi}\!\!\!L_{q} L_{Q}^{-1}\!\!   
\left[\Phi_{dc}+\frac{MV_{T}}{\omega L_{T}}\sin \tau \right]\!\cos ^{2}\tau\,d\tau, %
} %
\end{array}
\end{equation}
where $\xi_{0}\!=\!(1-\omega_{T}/\omega)$ is a detuning parameter
set to zero hereinafter since $\omega\approxeq\omega_T$ in
experiment. As seen from Eqs.\,(\ref{EQ08}), voltage-current
$V_{T}(I_{0})$ and voltage-flux (signal) $V_{T}(\Phi_{dc})$
characteristics of the SQUTRID are determined by the reciprocal
quantum inductance $L_{Q}^{-1}(\Phi_{e})$ averaged over a period of
oscillations in the tank. Due to the sharp dependence of
$L_{Q}^{-1}(\Phi_{e})$ in the vicinity of $\Phi_{e}=\Phi_{0}$ [see
Fig.\,3(d)], small variations of signal magnetic flux
$\delta\Phi_{dc}$ will lead to a substantial change in the reactive
part of the tank impedance and therefore in the $V_T(t)$ and
$\alpha_{T}(t)$ dependencies.

Equations (\ref{EQ08}) should be solved numerically because of the
strong nonlinearity of the $L_{Q}^{-1}(\Phi_{e})$ function, to which
the sought tank voltage amplitude $V_T$ enters through the external
flux $\Phi_{e}$. We are also interested in taking into account the
effect of noise (generally of complex nature and spectrum)
influencing the qutrit on the measured $V_{T}(I_{0})$ and
$V_{T}(\Phi_{dc})$ dependencies. To this end, a simplified model is
used in which the major part of the noise influencing the qutrit
loop is considered to be caused by the measurement circuit. The
noise from the circuit produces fluctuations of external flux
applied to the qutrit loop that change the qutrit quantum response
and, in turn, its back action to the $L_{T}C_{T}$ tank. %
If the inequality $\omega_{T}\ll\omega_{i}\ll\Delta E_{01}/\hbar$ is
valid for all the noise spectrum components $\omega_{i}$ affecting
the SQUTRID loop from the side of the measurement circuit, then the
task becomes easier, and the value $(L_{q}L_{Q}^{-1})_{eff}(f_{e})$
effectively contributing to the tank can be found using the method
of averaging over \textit{quasistationary} thermodynamic
fluctuations.\cite{15} In the Gauss-distributed noise approximation
we get
\begin{equation} \label{EQ09}
\displaystyle{ %
(L_{q}L_{Q}^{-1})_{eff}(f_{e})=\!
\frac{1}{\sigma\sqrt{2\pi}}\!\int\!\! df' e^{-f'^2\! / 2\sigma^2}
(L_{q}L_{Q}^{-1})(f_e+ f'),
}%
\end{equation}
where $\sigma =\sqrt{\left\langle \delta f_{e}^{2}\right\rangle }$ %
is the standard deviation of the noise flux associated with the
measurement circuit (and $\sigma^2$ is the noise flux variance).
Substituting this function, $(L_{q}L_{Q}^{-1})_{eff}(f_{e})$, of the
effective reciprocal normed to $L_q$ quantum inductance [see
Fig.\,3(d)] into Eq.\,(\ref{EQ08}), one can numerically obtain
voltage-current $V_{T}(I_{0})$ and voltage-flux $V_{T}(\Phi_{dc})$
characteristics of the SQUTRID that account for the
measurement-induced noise flux parametrized by its variance.


\section{Experimental results and discussion}

The main objective of the experimental design is to provide
conditions at which (i) degenerate levels of the separated wells
become split with $\Delta E_{01}\gg k_{B}T$ due to interwell
tunneling and (ii) all the frequencies in the spectrum of the
environment noise $\omega_{i}$ will be small compared to the rate of
transitions to the superposition level $E_{S1}$ and the lower
level $E_{-1}$ situated in the middle well, i.e., %
$\omega_{i}\ll\Delta E_{01}/\hbar;\,(E_{S0}-E_{-1})/\hbar\,.$ %
In order to meet these conditions, a 3D toroidal SQUTRID
loop\cite{12} with inductance $L_{q}=0.3$\,nH was made from pure
(99.999\%) niobium (Nb). The design of the 3D toroidal
superconducting loop (inset in Fig.\,1), which, in fact, becomes a
closed 3D cavity, favorably eliminates undesirable coupling of the
qutrit to the external electromagnetic environment. The only way for
electromagnetic interference to come in is through the narrow
(0.5\,mm) and long (8\,mm) channel made in the qutrit body for
wiring the coupling coil placed inside the qutrit toroidal cavity.
With these dimensions, such a channel acts as a below-cutoff
waveguide for all frequencies of
up to several hundred gigahertz providing very high attenuation. %
The coupling coil leads are properly filtered. Adjusted \textit{in
situ} atomic-size ScS contact was realized between a surface-cleaned
Nb needle and an annealed Nb plate with a crystallite size close to
0.5\,mm. A number of atoms $N$ in the contact opening can be
assessed as ratio of the contact critical current $I_c$ to its
quantizing value $(e\Delta_0/\hbar)$, giving $N\sim 10$. The sample,
the main part of the resonance tank, and the filter were cooled down
to $T=450$\,mK in a $^3$He pumped refrigerator cell (see Fig.\,1).
The first stage of the rf amplifier and additional filters were
placed at $T=1.5$\,K. A triple $\mu$-metal shield around the liquid
$^4$He Dewar and a superconducting Pb shield around the measuring
cell were used to reduce and stabilize the ambient magnetic field.

An unusually low resonance frequency $\omega_{T}/2\pi=5.79$\,MHz of
the $L_{T}C_{T}$ tank was chosen to decrease the potential variation
rate and to meet the adiabatic conditions for the qutrit quantum
dynamics even at high (such that $MI_{0}Q\sim\Phi_{0}$) amplitudes
of the rf generator current $I_{0}$. The calculated time ($\tau\sim
10^{-9}$\,s) of passing the dip of the function
$(L_{q}L_{Q}^{-1})_{eff}(f_{e})$ [Fig.\,3(d)] for this frequency
considerably exceeds the superposition setting time ($\tau_{S}\sim
3\!\times\!10^{-11}$\,s) in the three-well symmetrical potential
[Fig.\,3(a)]. The resonance tank permanently measures the state of
the quantum system which is weakly coupled to it and, at the same
time, generates additional noise in the superconducting loop. Thus,
we should expect that the measurement process limits the quantum
superposition in our system. However, detecting the averaged
curvature of the base energy level is still possible since the
uncertainty of the magnetic flux associated with the effect of the
measurement circuit and temperature is estimated to be as low as
$\sim\!10^{-2}\Phi _{0} $.

Figure.\,4(a) exhibits a set of experimental voltage-flux
characteristics $V_{T}(\Phi_{dc})$ of the SQUTRID with
$\beta_{L}\approx 4$ obtained while sweeping the external magnetic
flux $\Phi_{dc}$ for several rf generator current amplitudes $I_{0}$. %
Note that plateaus exist in the range around
$\Phi_{dc}=\Phi_{0}(n+1/2),$ shrinking with increasing $I_{0}$.
These plateaus correspond to a quasiautonomous $L_{T}C_{T}$ tank
with characteristic resonance impedance $R_c=V_T/I_0=13.3\,k\Omega$
and indicate that no inductance is inserted into the tank from the
qutrit that receive external flux
$\Phi_{e}(t)=\Phi_{dc}+\Phi_{ac}(t),\,
\Phi_{ac}(t)=MV(t)/(\omega_TL_T)$, while sweeping $\Phi_{dc}$ flux.
This, in turn, is evidence of the function
$(L_{q}L_{Q}^{-1})_{eff}(f_{e})$ becoming zero [see Fig.\,3(d),
curve\,2] and of a no-superposition qutrit state in the
respective range of $\Phi_{e}(t)$. %
The onset of nonlinearity in the $V_{T}(\Phi_{dc})$ curve is observed at %
$\Phi_{e}(t)=\Phi_{dc}+\Phi_{ac}(t)\approx n\Phi_{0}$, %
when the magnetic flux $\Phi_{e}(t)$ falls into the narrow region
around $n\Phi_{0}$ [Fig.\,3(d), curve\,2]
where the qutrit superposition nonlinearity is localized. %
Particularly, for the triangle-shaped signal curve [at
$I_0=2.55$\,nA in Fig.\,4(a)] with a plateau degenerated into the
point $(\Phi_{dc}=0.5\Phi_{0}, V_{T}=30\mu\text{V})$, %
we have $\Phi_{e}=\Phi_{dc}+\Phi_{ac} \approx\Phi _{0}$, %
where $\Phi_{ac}=MV_{T}/(\omega_{T}L_{T})\approx 0.5\Phi_0$ is the
amplitude of the \textit{ac} flux coming to the qutrit from the tank. %
As clearly seen in Fig.\,4a, the $V_{T}(\Phi_{dc})$ characteristics
of the tank are maximally affected by the quantum inductance
inserted from the qutrit at symmetry points $\Phi_{dc}=n\Phi_{0}$,
where the qutrit superposition nonlinearity is maximum. %
At low-to-moderate current amplitudes, $V_{T}(\Phi_{dc})$
dependencies are well described by the theoretical model [see
Fig.\,4(b)] that takes into account the noise influence of the
measuring channel, with independently measured SQUTRID parameters
and capacitance $C=4.36$\,fF being used.

\begin{figure}[t!]
\centering %
\includegraphics[width = 1.0\columnwidth]{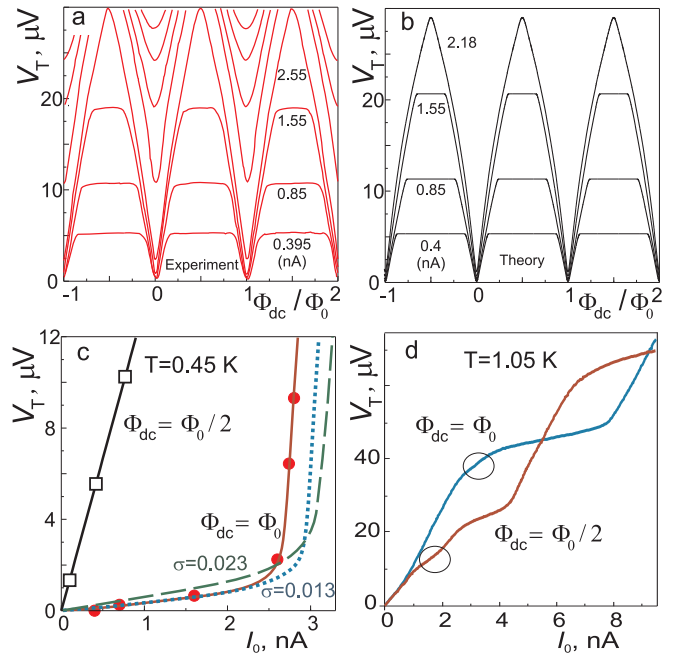}
\caption{\label{fig04} %
(Color online) SQUTRID characteristics: (a) experimental
voltage-flux (signal) characteristics $V_{T}(\Phi_{dc}/\Phi_{0})$
obtained for several amplitudes of rf generator current $I_{0}$
(indicated as the curve parameter) and (b) set of voltage-flux
characteristics calculated accounting for noise with standard
deviation $\sigma=0.013$ at currents $I_{0}$ close to the
experimental values (curve parameter). Independently measured
SQUTRID parameters $L_{q}=3\!\times\!10^{-10}$\,H, $\beta_{L}=4.0$,
$L_{T}=1.2\!\times\!10^{-6}$\,H, $M=1.52\!\times\!10^{-9}$\,H,
$Q=302$, $k^{2}Q=1.89$, and capacitance $C=4.36$\,fF were used in
the calculations. (c) Experimental (dots) and theoretical
voltage-current characteristics $V_{T}(I_{0})$ calculated for noise
flux standard deviations $\sigma=0.013$ (dotted line) and
$\sigma=0.023$ (dashed line) at $\Phi_{dc}=\Phi_{0}$ and
no-superposition experimental (squares) voltage-current
characteristic $V_{T}(I_{0} )$ at $\Phi_{dc}=\Phi_{0}/2$; (d)
smearing of $V_{T}(I_{0})$ curves when the refrigerator temperature
rises to $T=1.05$\,K. Nonlinearities due to the superposition
(circled) almost vanish.
}%
\end{figure}

Figure\,4(c) presents the initial parts of the SQUTRID
$V_{T}(I_{0})$ characteristics ($\beta _{L}\approx 4$) registered
for two values of the magnetic flux, $\Phi_{dc}=\Phi_{0}$ and
$\Phi_{dc}=\Phi_{0}/2$. The theoretical curves $V_{T}(I_{0})$ are
derived from Eqs.\,(\ref{EQ08}) with experimentally measured SQUTRID
parameters and averaging the superposition nonlinearity of the
quantum system over
low-frequency noise (\ref{EQ09}) with standard deviations %
$\sigma =0.013$ and $\sigma =0.023$. For the V-I curves with
$\Phi_{dc}=n\Phi_{0}$, the effective reciprocal quantum inductance
of the base superposition qutrit level, introduced into the tank at
low rf generator currents ($I_{0}\le 0.4$\,nA), leads to a large
shift in the tank resonance frequency from the generator frequency, %
which results in a decrease in the voltage detected at the
$L_{T}C_{T}$ tank down to a value comparable to the noise level. As
seen from Eqs. (\ref{EQ08}) and (\ref{EQ09}), at
$\Phi_{dc}=n\Phi_{0}$ and a
low current (linear) regime the voltage reads approximately as %
$V_{T}\approx\omega_{T}L_{T}I_{0}Q/|k^{2}Q(L_{q}L_{Q}^{-1})_{\!eff}^{\,\min }|$; %
that is, it is reduced by a factor of
$|k^{2}Q(L_{q}L_{Q}^{-1})_{\!eff}^{\,\min }|\gg 1$ compared to the
autonomous $L_{T}C_{T}$ tank voltage.

The theoretical curve $V_{T}(I_{0})$ with noise standard deviation
$\sigma=0.013$ coincides with the experimental curve in the range of
low-to-moderate generator currents. At higher $\sigma$, the factor
$|k^{2}Q(L_{q}L_{Q}^{-1})_{\!eff}^{\,\min }|$ decreases, and the
slope of the V-I curve in the area of small rf
generator currents becomes higher than in experiment %
[see the curve with $\sigma=0.023$ in Fig.\,4(c)]. At
$I_{0}>2.5$\,nA, the amplitude of \textit{ac} magnetic flux induced
in the SQUTRID loop $\Phi_{ac}=MV_T(I_{0})/(\omega_T L_T)$ exceeds
the half-width of the ``main dip'' of the effective reciprocal
quantum inductance, $\Phi_{ac}>0.02\Phi_0$ [see Fig.\,3(d)], and
then $V_{T}$ rapidly rises with the increase in the rf generator
current. Such behavior at high rf generator currents $I_{0}$
(nonlinear regime) is well described by the proposed theoretical
model. The observed quantitative deviation can be attributed to both
simplifications made when deriving the theoretical model (ideal ScS
contact, zero SQUTRID temperature, and Gaussian noise distribution)
and an experimental error in determining the SQUTRID parameters. %
The no-superposition quasiautonomous-tank experimental V-I curve at %
$\Phi_{dc}=\Phi_0/2$, which is related to the considered ``plateau
regime'' in $V_{T}(\Phi_{dc})$ characteristics, is shown in
Fig.\,4(c). It is seen that this V-I curve follows Ohm's law at the
specified $\beta_{L}\approx 4$. As discussed above, with a high
enough barrier, separating the two wells in the two-well potential,
the qutrit wave function remains localized in one of them during the
variation of the potential by the \textit{ac} component of the
external magnetic flux.

The impact of thermodynamic fluctuations on the effective reciprocal
quantum inductance (\ref{EQ09}) and on the rf V-I characteristics of
the SQUTRID is illustrated in Fig.\,4(d). The nonlinearity due to
the superposition of states is severely smeared out by temperature
at $T=1.05$\,K and becomes completely unobservable in our
experiments at helium bath temperature $T\ge 1.5$\,K. This result
can be explained by an increase in the higher-level population
because of the reduction of the ratio $\Delta E_{01}/k_B T \lesssim
1$ and an essential increase in the noise flux variance $\sigma^2$.

To see nonlinearities caused by the superposition of states in more
complicated potentials, we have increased the critical current of
the SQUTRID ScS contact up to $I_{c}\approx 8\,\mu$A. %
An increased $I_{c}$ and, thus $\beta_{L}$, parameter corresponds to
contacts with a greater number of atoms (atomic rows) at their
opening and hence decreased values of their
normal-state resistance $R_N$. %
Figure\,5(a) displays rf V-I characteristics of the SQUTRID with
$\beta_{L} \approx 7.3$ obtained at temperature $T=450$\,mK for
magnetic fluxes $\Phi_{dc}=n\Phi_{0}$ and
$\Phi_{dc}=(n+\!1/2)\,\Phi_{0}$. The initial parts of both branches
of the V-I curves remain almost linear
due to dynamic localization of the qutrit wave function. %
The effect of the reciprocal quantum inductance %
determined by the effective curvature (\ref{EQ09}) of the base
superposition level $E_{S0}$ is observable at generator currents
less than those needed to form ``classical'' steps in SQUID rf V-I
characteristics. Figure\,5(b) shows superposition energy levels
calculated for $\beta _{L}=7.3$ that appear in the four-well
symmetrical potential at $\Phi_{e}(t) =1.5\Phi_{0}$. Despite the
fact that, at chosen SQUTRID parameters, the environment needs to
absorb a
considerable amount of energy [$(E_{S0}-E_{-1})/k_{B}\approx 7$\,K] %
when the system relaxes to the lower level, the relaxation events
often occurred in the experiment. %
This can be caused by very high numbers of superposition energy
levels ($n=100$ in our case) that are less stable with respect to
noise and by possible dissipation effects related to an increase in
the number of atoms in the contact. %
The effect of the energy relaxation resulting in a jump to a linear
part of another ($\Phi_{dc}=0$) no-superposition branch of the V-I
curve is shown in Fig.\,5(a) by the up arrow. The reverse process of
the transition from the no-superposition V-I branch to the
superposition one shown by the down arrow reflects the formation of
the superposition state in the four-well symmetrical potential
configuration at a driving current that gives the total external
flux $\Phi_e(t)=1.5\Phi_0$.

\begin{figure}[t!]
\centering %
\includegraphics[width = 1.0\columnwidth]{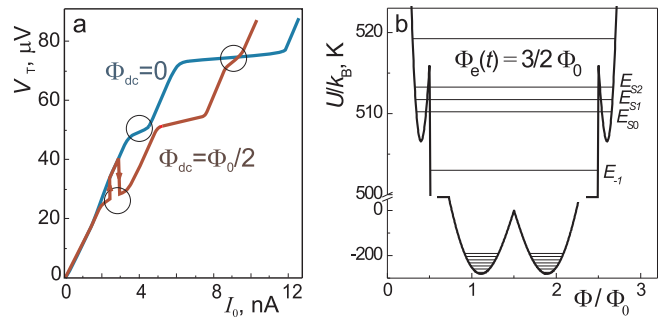}
\caption{\label{fig05} %
(Color online) Superposition of states in the superconducting ring
closed by ScS contact, with a four-well potential. (a) Experimental
voltage-current characteristics of the SQUTRID with
$\beta_{L}\approx 7.3$ obtained at $T=450$\,mK for external magnetic
fluxes $\Phi_{dc}=\Phi_{0}/2$ and $\Phi_{dc}=\Phi_{0}$. The areas
where the effect associated with maximal local curvature of the base
superposition level appears are circled. The jumps shown by the
arrows are due to relaxation events. (b) Potential energy and
superposition energy levels calculated for the SQUTRID parameters
taken from experimental data. Three superposition levels still form
in this four-well potential and are situated far above two middle
wells.
}%
\end{figure}

Note that small amplitudes of the SQUTRID driving current in quantum
measurements are the most favorable for quantum informatics.
The obtained conversion ratio %
($\eta = dV_{T}/d\Phi_{e}\approx 3\!\times\! 10^{10}$\,V$/$Wb) and
the SQUTRID performance in the three-well superposition (at
$\beta_{L}\approx 4$) can be further increased by increasing the
frequency $\omega$ while simultaneously and proportionally reducing
the generator current $I_{0}$. For example, at pumping frequency
$\omega/2\pi=100\,$MHz  we have $\eta$ of up to $5\times\! 10^{11}$\,V$/$Wb. %
In this case the SQUTRID voltage-flux characteristics would be
similar to those shown in Figs. 4(a) and 4(b) for small ($I_{0}\le
0.4$\,nA) current amplitudes. Since, as shown in
Ref.\,\onlinecite{16}, the detector and the tank temperature in
these measurements may be as low as $10$\,mK and the SQUTRID acts as
a sensor of parametric quantum inductance under adiabatic
conditions, such a device can be considered to be a representative
of the class of quantum-limited detectors.

\section{Summary and conclusions}

We have constructed and studied a new ideal parametric detector of
magnetic flux named rf SQUTRID which is based on quantum
superposition of three macroscopic flux states of a 3D toroidal
superconducting loop closed by Nb-Nb atomic-size point contact. %
Due to the specific form of the potential barrier, the small contact
capacitance, and the fast tunneling dynamics of formation of a
coherent three-well superposition state ($\tau_{S}\sim
3\!\times\!10^{-11}$\,s), the adiabatic conditions in the rf SQUTRID
are valid up to frequencies $\omega/2\pi\approx 3$\,GHz for small
driving amplitudes ($MI_0\sim 10^{-3}\Phi_0$). Therefore, the flux
detectors based on the dependence of the local curvature of the base
superposition energy level (or quantum inductance) on the external
magnetic flux at low temperatures fall well within the class of fast
and sensitive devices. Note that unlike the rf SQUTRID/SQUBID based
on the atomic-size point contact, similar detectors with contacts of
the SIS type will have slow tunneling dynamics of flux wave packets
and, respectively, a lower operating speed. The rf SQUTRID (as well
as rf SQUBID) is the flux magnetometer "dual" to the rf single
electron transistor (SET) electrometer,\cite{RF-SET} and so the main
applications of the rf SQUTRID will initially be experimental study
of physical processes in flux qubits and engineering readout systems
(sensors) for reading states in the small-scale superconducting
quantum registers using weak continuous measurements.

We stress the following results of this paper:
(i) The coherent superposition state $\Psi_{S0}(f)$ %
of wave functions corresponding to %
three distinct macroscopic states is formed in the three-well
symmetrical potential configuration at $\Phi_{e}=n\Phi_{0}$ in the
superconducting circuit with the clean atomic-size ScS contact under
study. %
(ii) The energy relaxation time $\tau_{\varepsilon}$ of the base
superposition state $\Psi_{S0}(f)$ to lower energy states can be
macroscopically large to perform quantum measurements. %
(iii) Unlike classical rf SQUID in a nonhysteretic regime, the
nonlinear properties of the superposition in the three-well
potential under the condition %
$\omega/2\pi\ll \nu_{01}\ll\Delta_{0}/h $ enable making a sensitive
and fast parametric detector of magnetic flux without a
quasiparticle current, i.e., principally a nondissipative rf SQUTRID
device. %
(iv) An unusually large splitting $\Delta E_{01}/k_{B} = 1.5 - 2$\,K 
and a very large (at low noise variance) $L_{q}L_{Q}^{-1} \approx
200$, specifying the nonlinearity of the studied quantum system,
make the rf SQUTRID with a ScS contact a very promising element for
quantum informatics.

\acknowledgments %
The authors would like to thank V.A. Khlus and G.M. Tsoi for helpful
discussions. V.\,I.\,Sh. acknowledges partial support from NAS
Ukraine through Project. No. 4/11 NANO.

\end{document}